\documentclass[twocolumn,superscriptaddress,amsmath,amssymb,aps, pre]{revtex4-1}

\usepackage[utf8]{inputenc}
\usepackage{subfigure}
\usepackage{graphicx}
\usepackage{dcolumn}
\usepackage{bm}
\usepackage{enumitem}
\usepackage{xcolor}

\begin{document}

\title{The saddle-straddle method to test for Wada basins}

\author{Alexandre~Wagemakers}
\email{Corresponding author: alexandre.wagemakers@urjc.es}
\affiliation{Nonlinear Dynamics, Chaos and Complex Systems Group, Departamento de  F\'isica, Universidad Rey Juan Carlos\\ Tulip\'an s/n, 28933 M\'ostoles, Madrid, Spain, ORCID:  0000-0003-2428-5744}

\author{Alvar~Daza}
 \affiliation{Nonlinear Dynamics, Chaos and Complex Systems Group, Departamento de  F\'isica, Universidad Rey Juan Carlos\\ Tulip\'an s/n, 28933 M\'ostoles, Madrid, Spain, ORCID: 0000-0002-8299-0196}

\author{Miguel A.F. Sanju\'{a}n}
\affiliation{Nonlinear Dynamics, Chaos and Complex Systems Group, Departamento de  F\'isica, Universidad Rey Juan Carlos\\ Tulip\'an s/n, 28933 M\'ostoles, Madrid, Spain}
\affiliation{Department of Applied Informatics, Kaunas University of Technology, Studentu 50-415, Kaunas LT-51368, Lithuania}
\date{\today}

\begin{abstract}
First conceived as a topological construction, Wada basins abound in dynamical systems. Basins of attraction showing the Wada property possess the particular feature that any small perturbation of an initial condition lying on the boundary can lead the system to any of its possible outcomes. The saddle-straddle method, described here, is a new method to identify the Wada property in a dynamical system based on the computation of its chaotic saddle in the fractalized phase space. It consists of finding the chaotic saddle embedded in the boundary between the basin of one attractor and the remaining basins of attraction by using the saddle-straddle algorithm. The simple observation that the chaotic saddle is the same for all the combinations of basins is sufficient to prove that the boundary has the Wada property.

\end{abstract}

\maketitle

\section{\label{sec:Introduction}Introduction}
In 1917, Kunizo Yoneyama published a work on topology where he described how to divide a region of the plane in three or more connected sets sharing a common boundary \cite{yoneyama_theory_1917}. He attributed the authorship of the procedure to his advisor Takeo Wada, and since then these regions were called Wada lakes. At first, the intriguing properties of the Wada lakes were studied within a topological context \cite{hocking1988topology}. For example, the Polish topologist Kazimierz Kuratowski showed that if a boundary separates at the same time three or more connected regions in the plane, then the boundary must be an indecomposable continuum \cite{kuratowski_sur_1924, sanjuan1997indecomposable}. Years later, Wada lakes were studied by James Yorke and collaborators under the perspective of dynamical systems \cite{kennedy_basins_1991, nusse_wada_1996}. They studied the set of initial conditions leading to a particular attractor, called the basins of attraction, in a forced damped pendulum. They demonstrated numerically that for a particular set of parameters, the forced damped pendulum presents three basins of attraction sharing the same boundary, that is, they are Wada basins. The Nusse-Yorke condition to assert the Wada property in \cite{nusse_wada_1996} was based on the computation of the unstable manifold of a saddle point, which intersected all the three basins. This is how an apparently inconceivable geometry arose in such a simple system as the forced damped pendulum. The cumbersome structure of the Wada basins implies a particular kind of unpredictability, since a small perturbation in the initial conditions lying on a Wada boundary may lead the system to any of the system's attractors. This is one of the reasons that explain why the Wada property has been so intensively studied in dynamical systems.

Since the pioneering works of Yorke and collaborators \cite{kennedy_basins_1991, nusse_saddle-node_1995, nusse_wada_1996, nusse_fractal_2000}, the Wada property has been found in many different cases: chaotic scattering \cite{poon_wada_1996}, Hamiltonian systems \cite{aguirre_wada_2001}, fluid dynamics \cite{toroczkai_wada_1997}, delayed systems \cite{daza_wada_2017}, etc. In most of these works, the authors used the Nusse-Yorke condition mentioned earlier. However, Daza et al. \cite{daza_testing_2015, daza_ascertaining_2018} have recently proposed two new methods to test for the Wada property. The first one was the \textit{Grid method} \cite{daza_testing_2015}, which was based on successive refinements of a grid that permit to check if every boundary point of the initial grid is a Wada point. This method allows one to determine the presence of Wada basins in any kind of dynamical system including systems that show disconnected Wada basins and even partially Wada basins \cite{zhang_wada_2013}. Besides, it provides a parameter $0\leq W\leq 1$ to characterize the Wada property that gives account of the number of Wada points divided by the total number of boundary points. The second one receives the name of \textit{Merging method} \cite{daza_ascertaining_2018}, and is based on an equivalent definition of the Wada property: three or more basins have the Wada property if their boundary remains unaltered when all but one basin are merged. Given a basin of attraction, this method can determine whether a basin is Wada in just a few seconds. { However the computation of the basin of attraction can be a long process depending on the system and the required resolution of the grid}. The precision of the result is determined by the resolution of the basin and a \textit{fattening-parameter} associated to the merging method.

Here, we propose a new approach to check the Wada property. The method explained here makes use of the merging property of the Wada basins described in \cite{daza_ascertaining_2018}. In essence, the new technique verifies whether there is a unique set, called the chaotic saddle, which is invariant under the operation of merging several basins of attraction together. If indeed the chaotic saddle embedded in the boundary is unique, the basins have the Wada property.

Although this method works only for connected Wada basins and requires detailed knowledge of the dynamics of the system, it is fast and, furthermore, it is able to search for the Wada property up to any desired precision.

\section{\label{sec:chaotic_saddles}Chaotic saddles and Wada basins}

Connected Wada basins are separated by a unique connected boundary \citep{kennedy_basins_1991}. In terms of the dynamics, this means that there is a unique invariant set under forward iteration, i.e., there is only one stable manifold. The existence of a unique stable manifold involves the existence of a unique saddle. This saddle must be an indecomposable continuum, as shown by Kuratowski \cite{kuratowski_sur_1924}. Therefore, we can argue that connected Wada basins do happen in systems with three or more possible outcomes and only one saddle, which must be a chaotic saddle. The main goal is to construct a numerical proof showing that there is a unique chaotic saddle in the phase space, what would prove the basins to be Wada.

We can relate the previous approach to Wada basins with the merging property described in~\cite{daza_ascertaining_2018}. The merging property basically says that, given $N_A \geq 3$ basins of attraction, and being their boundaries $\partial B_i$ defined as the boundary between a basin $B_i$ and all the other merged basins $\bigcup\limits_{j\neq i} B_j$, they will possess the Wada property if and only if $\partial B_i=\partial B_j$,  $\forall i \neq j, i=1,\ldots, N_A$.

\begin{figure}
\begin{center}
\includegraphics[width=0.45\textwidth]{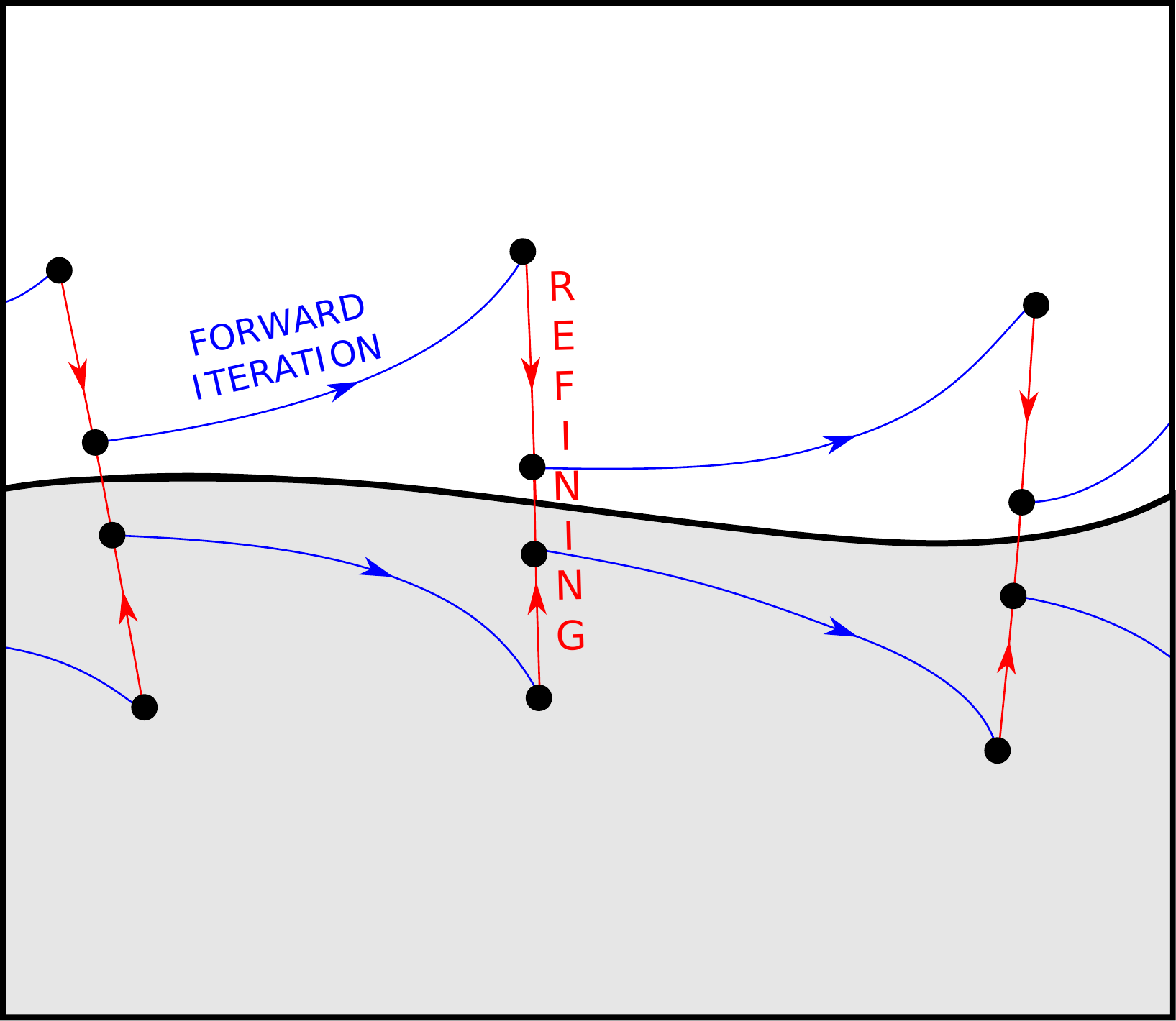}
\end{center}
\caption{\label{fig:ss_method} \textbf{Sketch of the saddle-straddle algorithm.} Initially, two points are selected in such a way that each one lies on a different basin. Then, a bisection method is applied to reduce the distance between the two points to a desired accuracy. After that, the resulting points are iterated and the segment expands, so that the process must start over again. As a result, we obtain a set of arbitrarily small segments straddling the saddle.}
\end{figure}

The merging property of Wada basins and the previous observation about chaotic saddles and Wada boundaries can be connected through the {\it saddle-straddle algorithm}. The chaotic saddle is a special case of a limit set on the boundary named {\it basic set} \cite{grebogi1987, grebogi1988}, that can be approximated with the saddle-straddle algorithm \cite{battelino1988multiple,nusse2012dynamics}. This is a computational technique that allows to find an arbitrarily accurate set of segments belonging to the saddle. The algorithm starts with two initial conditions, each one lying on a different basin. The method receives its name because the segment is straddling the boundary. Firstly, by using a bisection routine, the saddle-straddle algorithm refines the original segment up to a given resolution. { In our implementation, its size is no longer than $1\cdot 10^{-8}$}. Then, the extreme points of the segment are iterated forward under the dynamics of the system. The segment expands because of this forward evolution in the vicinity of a unstable manifold. Thus, the segment must be refined again to recover the desired accuracy. Then, the process starts over. The whole procedure can be visualized in Fig.~\ref{fig:ss_method}. When the desired number of segments composing the saddle is reached, the saddle-straddle algorithm stops. Therefore, we end with an arbitrarily accurate picture of the saddle.

As explained above, the saddle-straddle algorithm uses two initial conditions lying on different basins. However, Wada basins only happen for three or more attractors, so we have to be careful about how to apply the saddle-straddle algorithm to Wada basins. Fortunately, the merging property indicates how to proceed. We can apply the saddle-straddle algorithm to every basin $B_i$ and the basin formed by merging the remainder $\bigcup\limits_{j\neq i} B_j$. In the case that the basins have the Wada property, the chaotic saddles obtained by applying the saddle-straddle algorithm to the different combinations of merged basins must coincide. In the next sections, we explain how to implement all these steps into a single algorithm and we illustrate it by using the paradigmatic examples of the damped forced pendulum and the H\'enon-Heiles Hamiltonian.


\section{\label{sec:ss_method}Description of the saddle-straddle method to test for Wada basins}

The saddle-straddle algorithm has been designed to get an accurate picture of a chaotic saddle that lies in a boundary. This boundary separates at least two basins, so the first step is to correctly identify all the attractors in the phase space to tag without ambiguity the basins. Notice that there is no need to compute the basin of attraction on a finite grid, but we just need to find suitable initial conditions leading to the different attractors. For each of the $N_A$ attractors, we will define a pair of basins formed by the basin $B_i$ associated to the attractor $i$, and another basin $M_i=\bigcup\limits_{j\neq i} B_j$, which is the result of merging the basins of all the remaining attractors. In this way, we obtain $N_A$ different pairs of basins $(B_i, M_i)$.

The saddle-straddle algorithm computes very short segments that straddle the basins $B_i$ and $M_i$ with a desired tolerance. In the following we will use the term \textit{algorithm} to refer to the way of computing the saddles and the term \textit{method} for verifying the Wada property. For each pair of basins the algorithm computes $n_p$ segments very close to the saddle corresponding to the boundary between the basins $B_i$ and $M_i$. Based on the previous arguments, we can argue that if the computed saddles are the same for all the pairs of basins, then this means that there is a unique boundary and, as a consequence, the basins have the Wada property.

\begin{figure*}
\begin{center}
\includegraphics[]{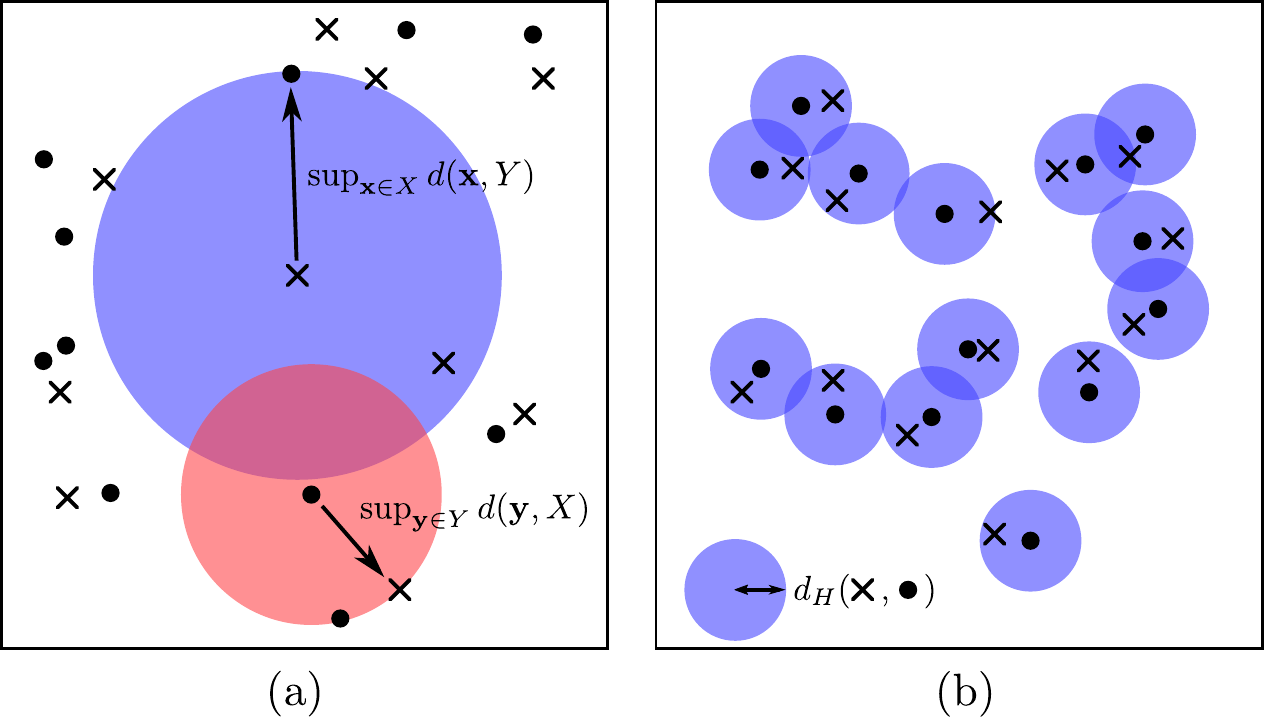}
\end{center}
\caption{\label{pic_hausdorff} \textbf{Interpretation of the Hausdorff distance.} In (a), a blue disk represents the longest distance that one would have to travel to go from the set of the crosses to the set of dots, while the red disk is the largest distance from the set of dots to the set of crosses. In (b) we represent an example of a set of dots covered with disks with the radius $d_H$, the Hausdorff distance between the sets of crosses and dots. The two sets fit in the surface defined by the union of the disks.}
\end{figure*}

Even though the chaotic saddle of a dynamical system may be unique, its precise numerical construction depends on the chosen initial conditions of the saddle-straddle algorithm. For each selected pair of basins, we may obtain different orbits embedded in the same saddle. Thus, it is not trivial to compare two orbits corresponding to the same chaotic set with different initial conditions. The chaotic orbits will be very similar and will share common properties. However, they will never coincide exactly in the phase space. To solve this difficulty, we propose to measure the distance between two sets of points using the Hausdorff distance \cite{edgar2007measure}. This distance measures the longest distance that we can possibly travel to get from one set to another as shown in Fig.~\ref{pic_hausdorff} (a), where the largest distances between one set and another are represented with disks. In other words, it guarantees that every point in one set is within a distance $d_H$ of a point of the other set. In the Fig.~\ref{pic_hausdorff} (b) the Hausdorff distance between the set of crosses and the set of bullets is represented by a blue disk of radius $d_H(\times,\bullet)$. We have drawn this disk around each bullet, and as a consequence of the definition of the Hausdorff distance, every cross of the other set fits within one of the disks.
If we define the distance between a single point $\bf x$ and the set $X$ as:
\begin{equation}
  d({\bf x},X) =\min_{{\bf y} \in X}(|| {\bf x} - {\bf y}||),
\end{equation}
the Hausdorff distance results:
\begin{equation}
d_H(X,Y) = \max\{\sup_{{\bf x} \in X} d({\bf x},Y); \sup_{{\bf y} \in Y} d({\bf y},X)\}.
\end{equation}
Computing this distance involves finding the closest points from one set to each other. Fortunately, there are efficient algorithms to find the nearest neighbors between two large sets of points such as the $k$-$d$ tree algorithm \cite{Friedman_kdtree}. The results of the saddle-straddle computation are compared pairwise to the other results of the different basin combinations $(B_i, M_i)$.

After computing the Hausdorff distance a question arises: is this a sufficiently small distance to consider the sets identical? To answer this question, we must first define the diameter of a set
\begin{equation}
d_s(A) = \sup\{ ||{\bf x} - {\bf y}  || :  {\bf x},{\bf y} \in A \}.
\end{equation}
Simply put, it is the largest Euclidean distance between two points of an attractor. This allows us to define the following criterion: {\it if the measured Hausdorff distance between the sets is small with respect to the diameter $d_s$ of one of the set, we can say that the sets correspond to the same saddle.}

We can summarize the steps of the method as follows:
\begin{enumerate}
  \item First, we classify the attractors of the dynamical system and we assign an integer $i$ to each basin.
  \item We form the pairs of basins as follow: for each attractor, we define the basin $B_i$ of the attractor and the basin $M_i$ as the union of the remaining basins. There are as many pairs of basins as attractors.
  \item We compute the saddle for each pair of basins using the saddle-straddle algorithm.
  \item The saddles are compared pairwise using the Hausdorff distance $d_H$. We consider that the saddles belong to the same set when the distance $d_H$ is small compared to the diameter of the set $d_s$. {In case the sets have different diameters, we will pick the largest.}
  \item If all the previous comparisons are successful, then there is only one boundary and the basins of attraction possess the Wada property.
\end{enumerate}

{ It is possible to discard directly the hypothesis of a Wada basins if one of the Hausdorff distances is about the diameter of the set. This generally corresponds to the case of disjoint sets. Another common negative result arises when a saddle point is present on a smooth boundary. Its diameter, very small, is comparable to the straddle segment size. This is an evidence pointing out that we are not in the presence of a chaotic saddle embedded in the boundary.}

An important control parameter of the method is the number of points $n_p$ that the saddle-straddle algorithm should compute for each pair of basins. If the number of points is too small, we may not compute correctly $d_H$ since the set of points is not representative of the attractor. We will give a few examples of the application of the algorithm with well-known examples of Wada and partially Wada basins.

\section{\label{sec:forced_pendulum}Application of the saddle-straddle method to the forced damped pendulum with Wada basins}

The forced pendulum is an appropriate example to try the algorithm. Depending on the parameters of the system, its basins of attraction may have the Wada or the partially Wada property. Partially Wada basins refer to a situation with boundaries composed of both Wada and non-Wada points \cite{zhang_wada_2013, Aguirre2009}.

\begin{figure*}
\begin{center}
\subfigure[]{\includegraphics[width=0.35\textwidth]{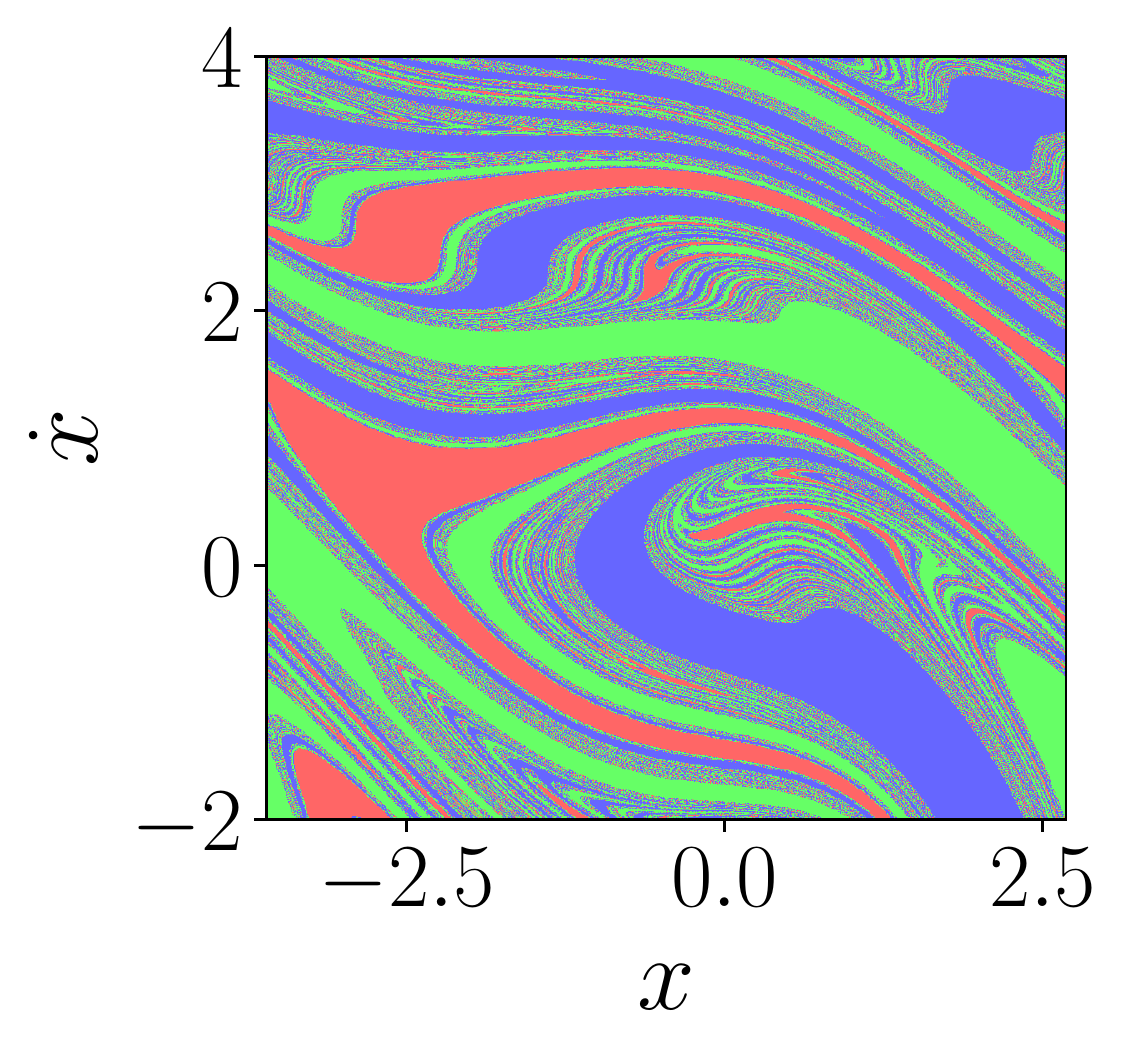}}
\subfigure[]{\includegraphics[width=0.35\textwidth]{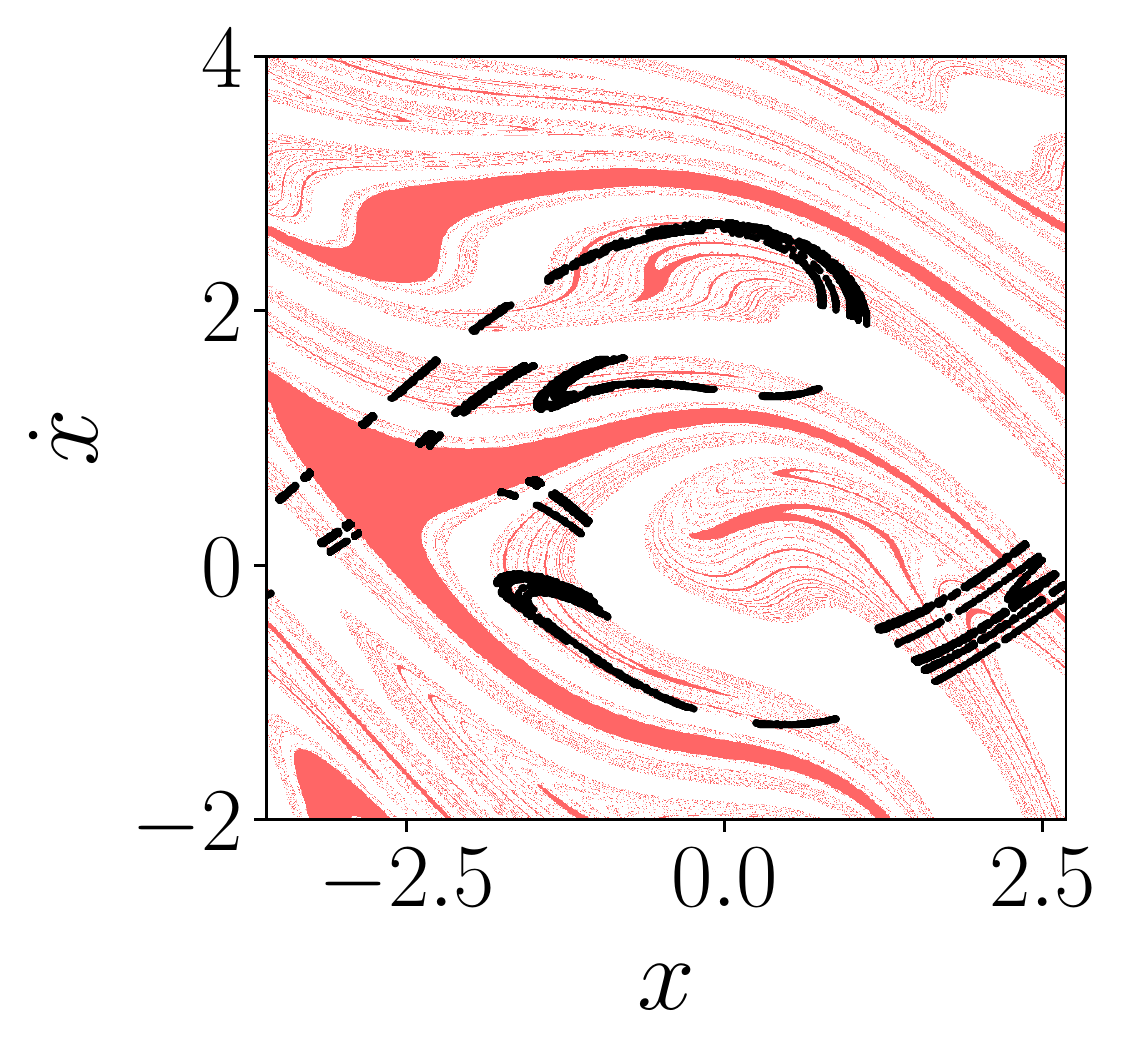}}
\subfigure[]{\includegraphics[width=0.35\textwidth]{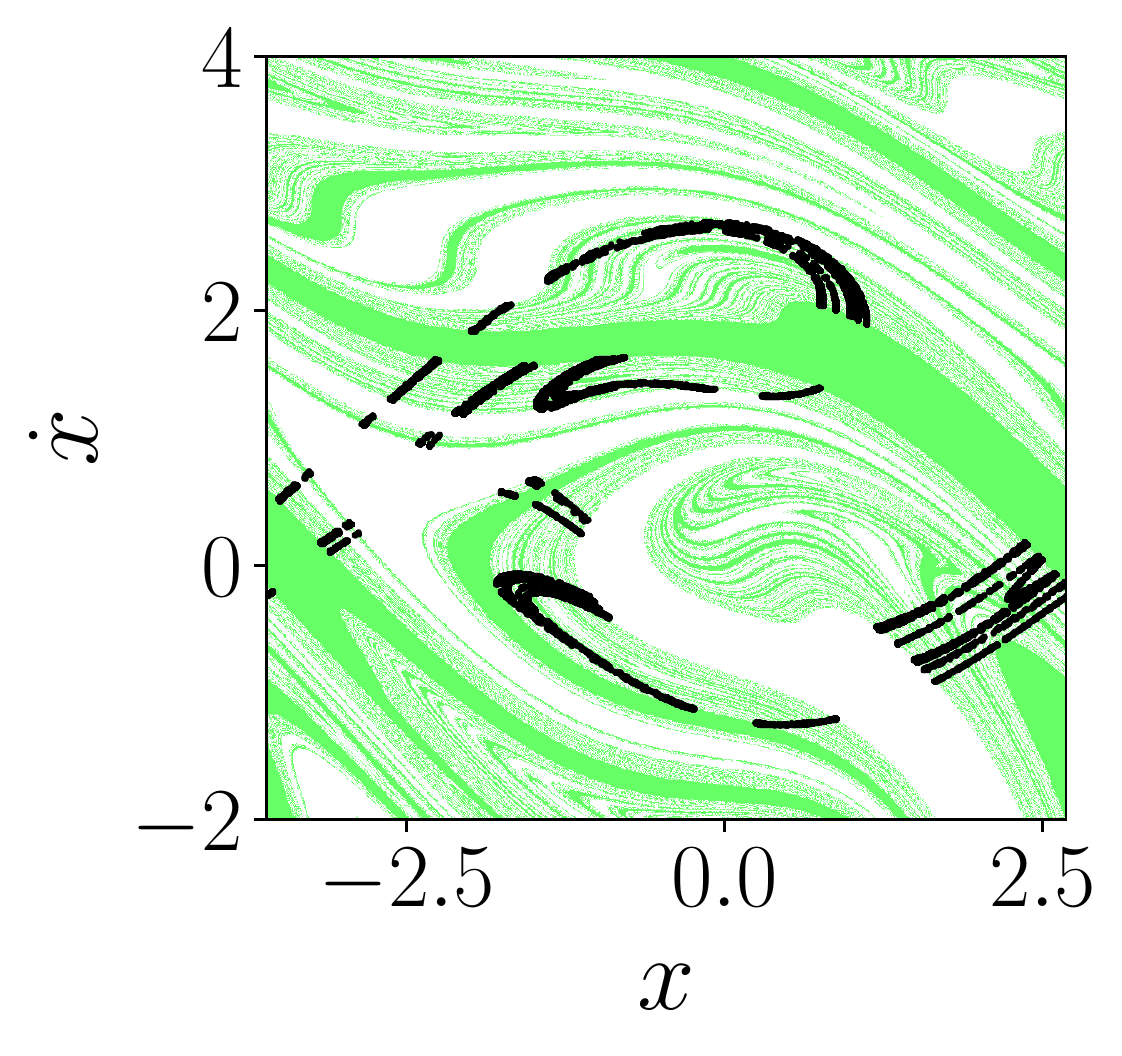}}
\subfigure[]{\includegraphics[width=0.35\textwidth]{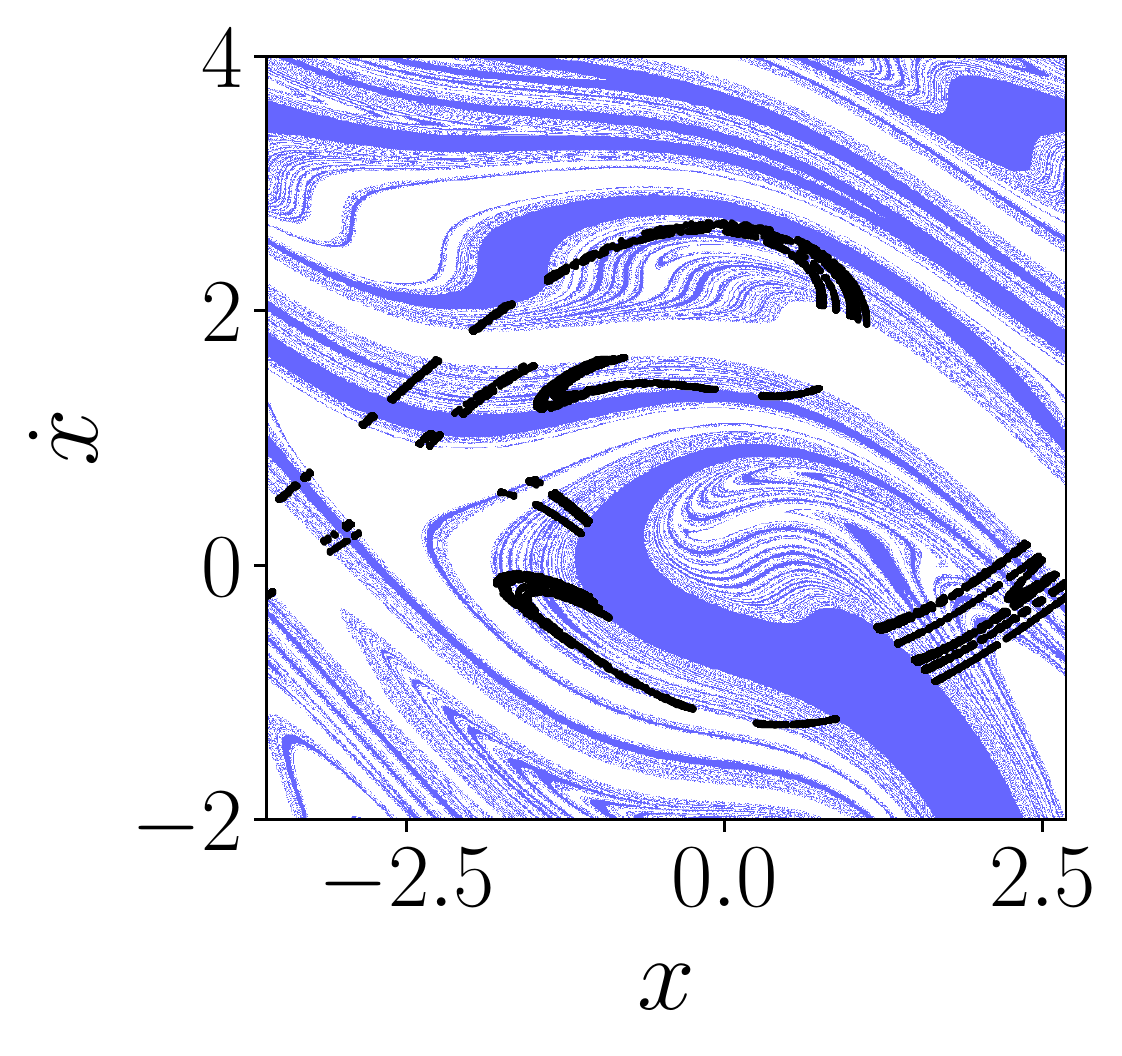}}
\end{center}
\caption{\label{ss_wada_pendulum} \textbf{Saddle-straddle trajectories for $\ddot{x}+0.2\dot{x}+\sin x=1.666\cos t$} This is a typical example of a fractal Wada basin. (a) Original basin of attraction. (b) Saddle computed with the pair of basins formed by the red basin and the green and blue merged together. The merged basins are displayed in white. (c) Green basin against the red and blue merged together. (d) Blue basin against the red and green merged together. The points of the computed saddle-straddle trajectories are superimposed over the basin with black dots.}
\end{figure*}

For the Wada case in Fig.~\ref{ss_wada_pendulum} (a), we have three attractors whose basins share the same boundary. Panels (b), (c) and (d) of Fig.~\ref{ss_wada_pendulum} show the chaotic saddles in the phase space (black dots) over the merged basins. The basin corresponding to the attractor is filled with the original color of Fig.~\ref{ss_wada_pendulum} (a) and the merged basins are displayed in white. A visual inspection suggests that the computed chaotic saddles are the same. To confirm this first impression, we compute the Hausdorff distances between each set. For simplicity, we refer to the sets of points representing the saddles with the letters $r,g,b$ for the colors red, green or blue corresponding to the basins of the Fig. \ref{ss_wada_pendulum} (b) (c) and (d) respectively.
The results of the comparisons for 40000 points are: $d_H(r,g) = 0.04686$, $d_H(r,b) = 0.04689$ and $d_H(b,g) = 0.04650$. The distances $d_H$ are very small compared to the diameter of the saddle under study measured as $d_s(r)\simeq 2\pi$, which confirms our first impression that all sets of points belong to the same saddle.

\begin{figure*}
\begin{center}
\subfigure[]{\includegraphics[width=0.35\textwidth]{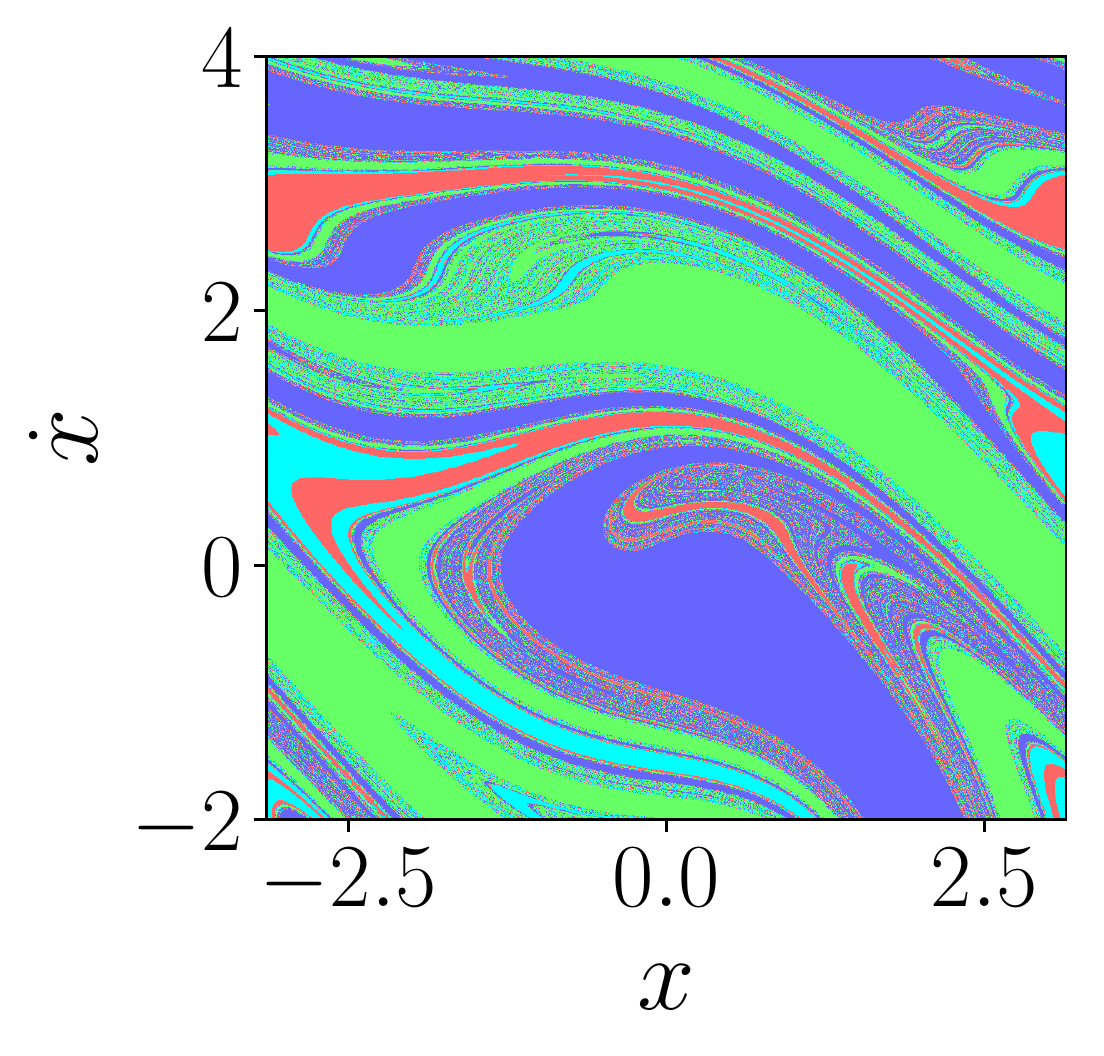}}
\subfigure[]{\includegraphics[width=0.35\textwidth]{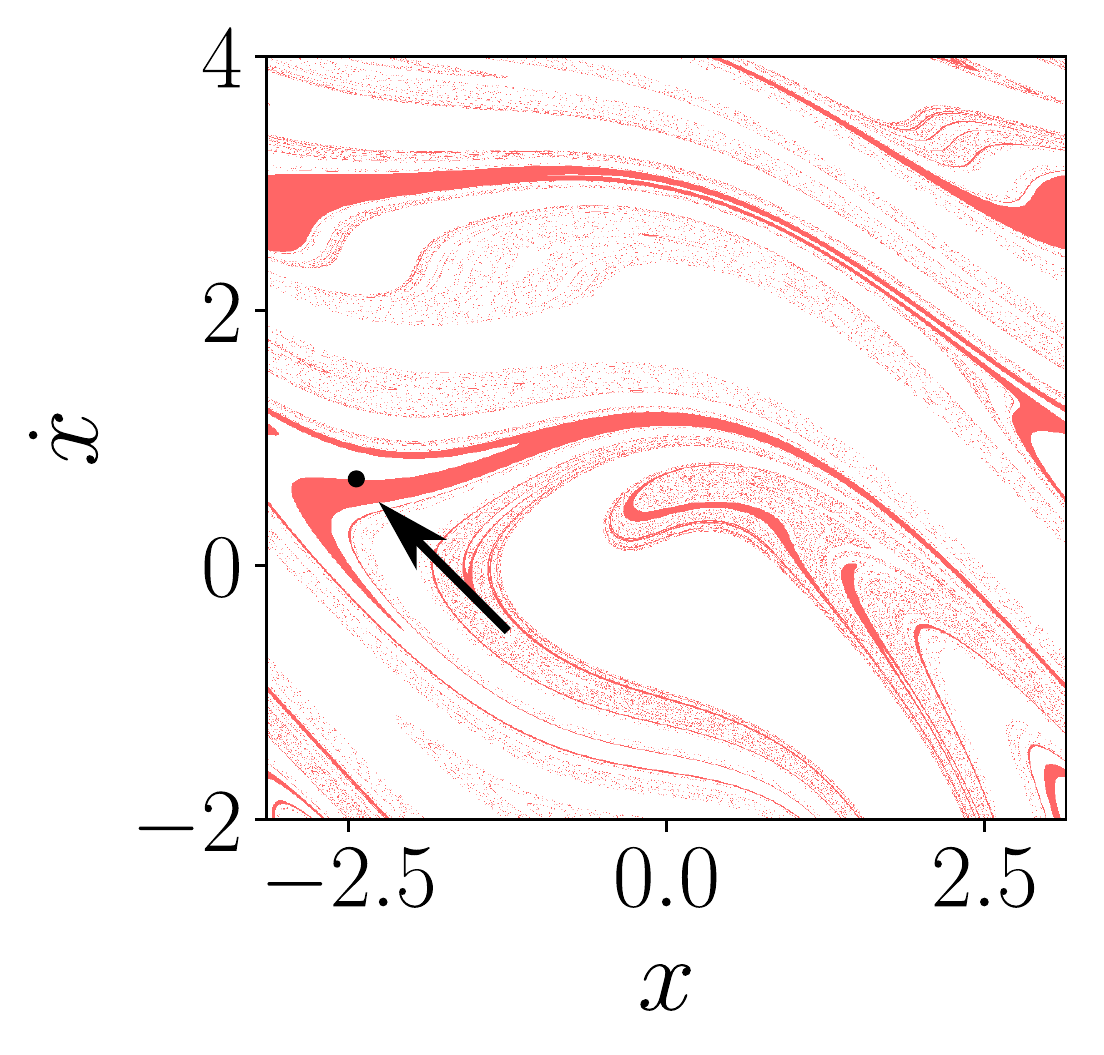}}
\subfigure[]{\includegraphics[width=0.35\textwidth]{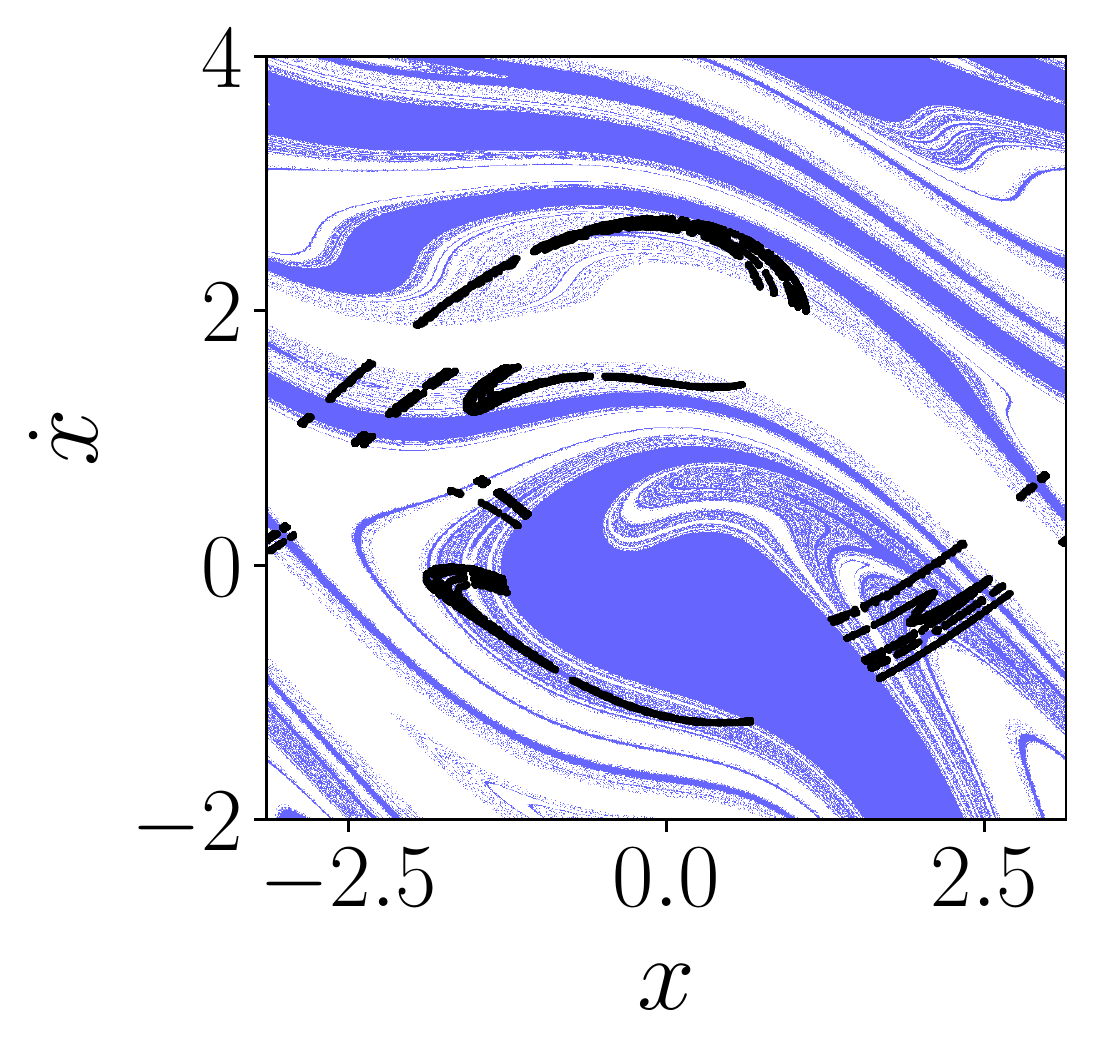}}
\subfigure[]{\includegraphics[width=0.35\textwidth]{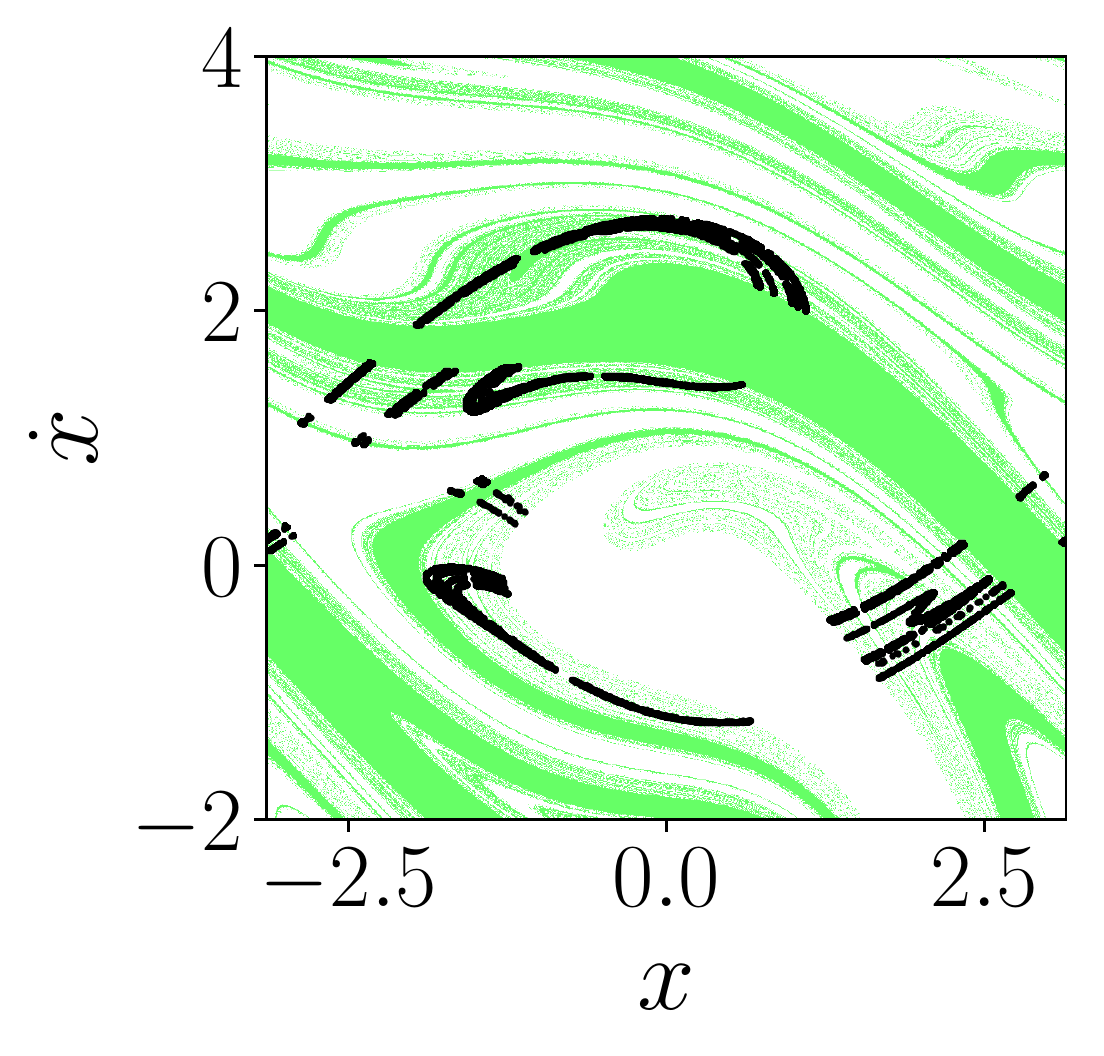}}
\subfigure[]{\includegraphics[width=0.35\textwidth]{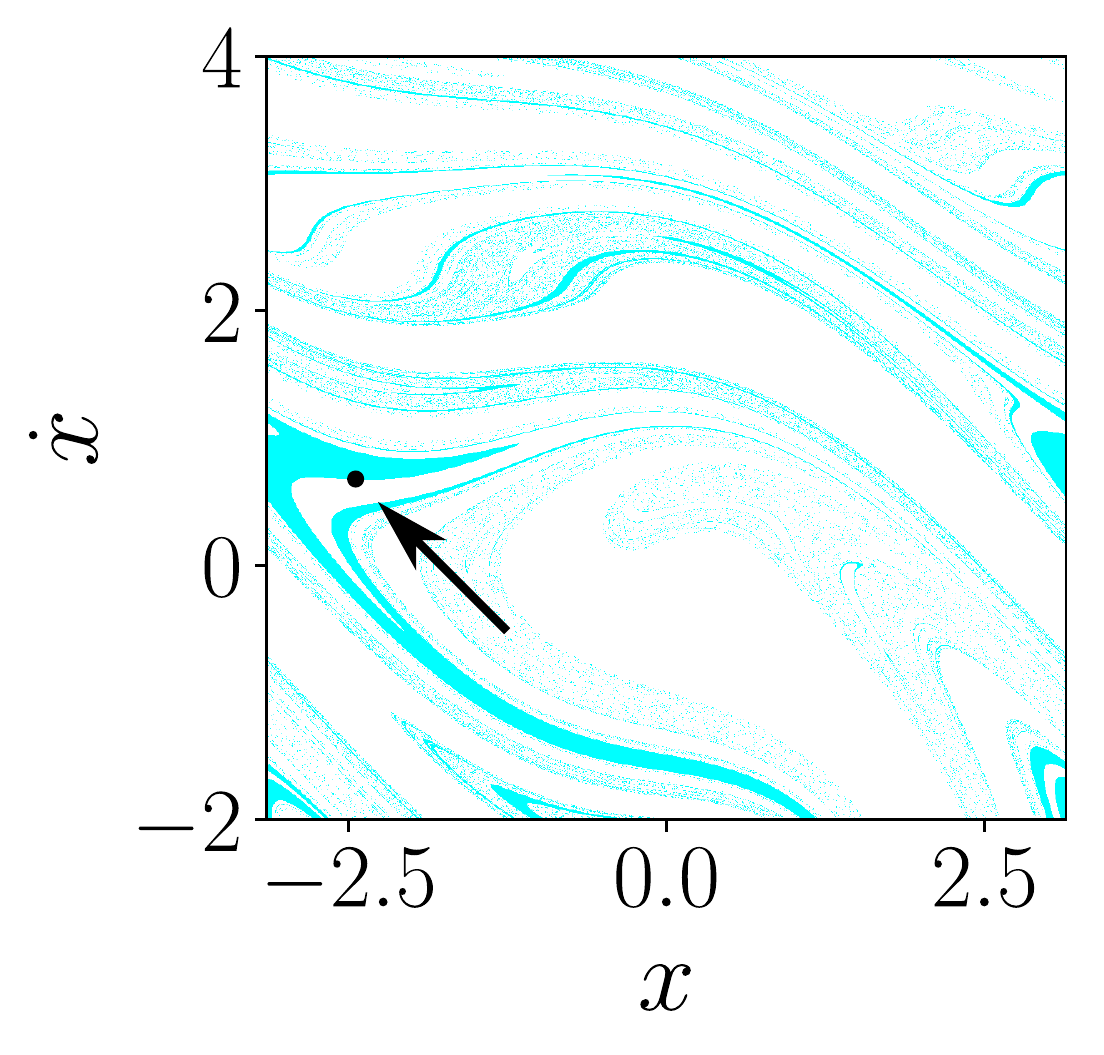}}
\end{center}
\caption{\label{ss_partial_pendulum} \textbf{Saddle-straddle trajectories for $\ddot{x}+0.2\dot{x}+\sin x=1.71\cos t$.} The basins of the forced pendulum with these parameters have the partial Wada property. (a) Original basins of attraction. (b) Saddle computed with the pair of basins formed by the red basin and the other three basins merged together. The merged basins are displayed in white. (c) Green basin against the other three basins  merged together. (d) Blue basin against the other three basins merged together. (d) Cyan basin against the other three basins merged together. The points of the computed saddle-straddle trajectories are superimposed over the basin with black dots.}
\end{figure*}

To contrast with this successful example, we expose a case of a partially Wada basin with four attractors presenting smooth and fractal boundaries. Figure~\ref{ss_partial_pendulum} shows the result of the saddle-straddle algorithm applied to the combination of basins described earlier for the system with four attractors. It appears clearly that two sets of points are very similar (Fig.~\ref{ss_partial_pendulum} (c) and (d)) whereas the other two are reduced to a single point marked with an arrow for clarity.

We will refer to the sets of points with the indices $r,g,b$ and $c$ that correspond to the color red, green, blue and cyan associated to the attractors defined in the basins of the Fig. \ref{ss_partial_pendulum} (b), (c), (d) and (e). We need to perform six pairwise comparisons between the sets. The Hausdorff distances computed between each pair of sets for 40000 points show clearly that there is not a unique boundary: $d_H(r,g) = 5.604$, $d_H(r,b) = 5.604$, $d_H(g,c) = 5.604$, $d_H(b,c) = 5.604$, $d_H(r,c) = 5.02\cdot 10^{-9} $ and $d_H(g,b) = 0.064$.
The very small distance $d_H(r,c)$ shows that the saddles computed from the merged basin represented in Figs.~\ref{ss_partial_pendulum} (b) and (e) are identical. { The values of the diameters $d_s(r)$ and $d_s(c)$ fall bellow the precision of the algorithm $1\cdot 10^{-8}$ meaning that we are in presence of a saddle point with a smooth boundary in phase space. With this observation in hand, we can conclude that the basin cannot have the Wada property.} The algorithm also finds a small distance between the saddles of the merged basin $g$ and basin $b$. This is the fractal structure depicted in Figs.~ \ref{ss_partial_pendulum} (c) and (d) with a diameter $d_s(b)=d_s(g)=2\pi$. We can conclude that there are two different saddles coexisting in phase space that correspond to different boundaries. When this kind of situation occurs, we can only affirm that the basin has at best the partially Wada property.

\section{\label{sec:HH}Application of the saddle-straddle method to the H\'enon-Heiles Hamiltonian}

\begin{figure*}
\begin{center}
\subfigure[]{\includegraphics[width=0.35\textwidth]{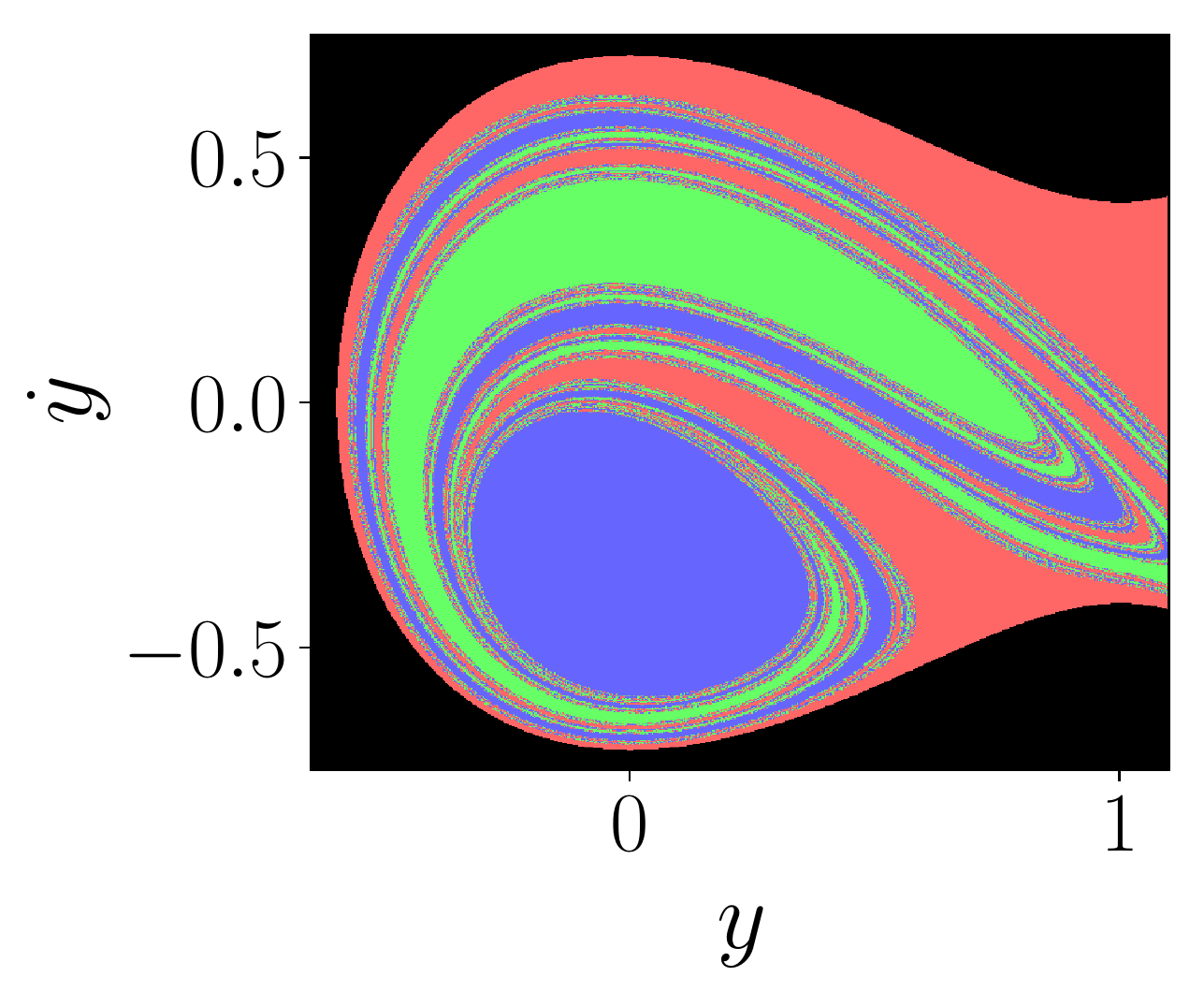}}
\subfigure[]{\includegraphics[width=0.35\textwidth]{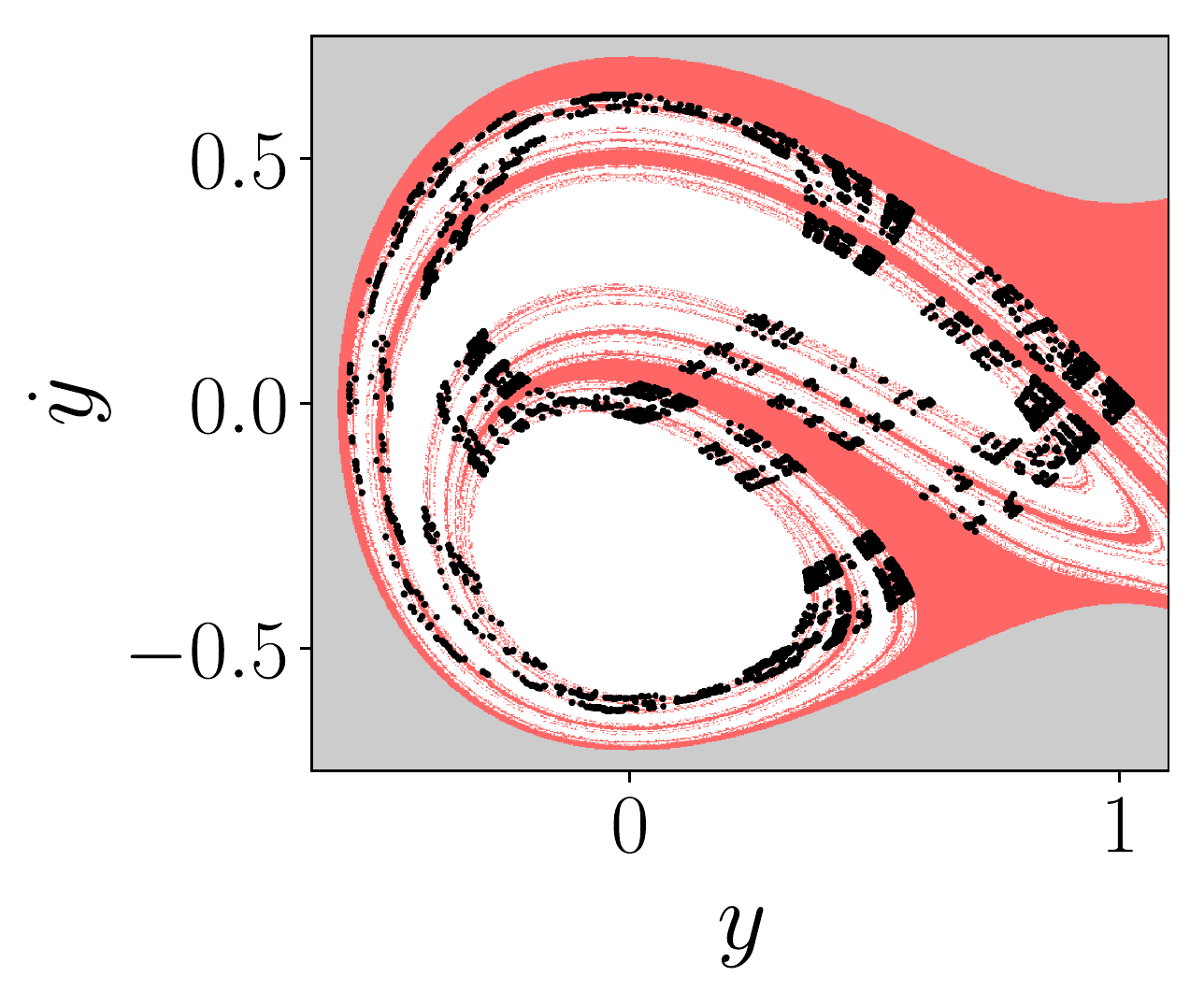}}
\subfigure[]{\includegraphics[width=0.35\textwidth]{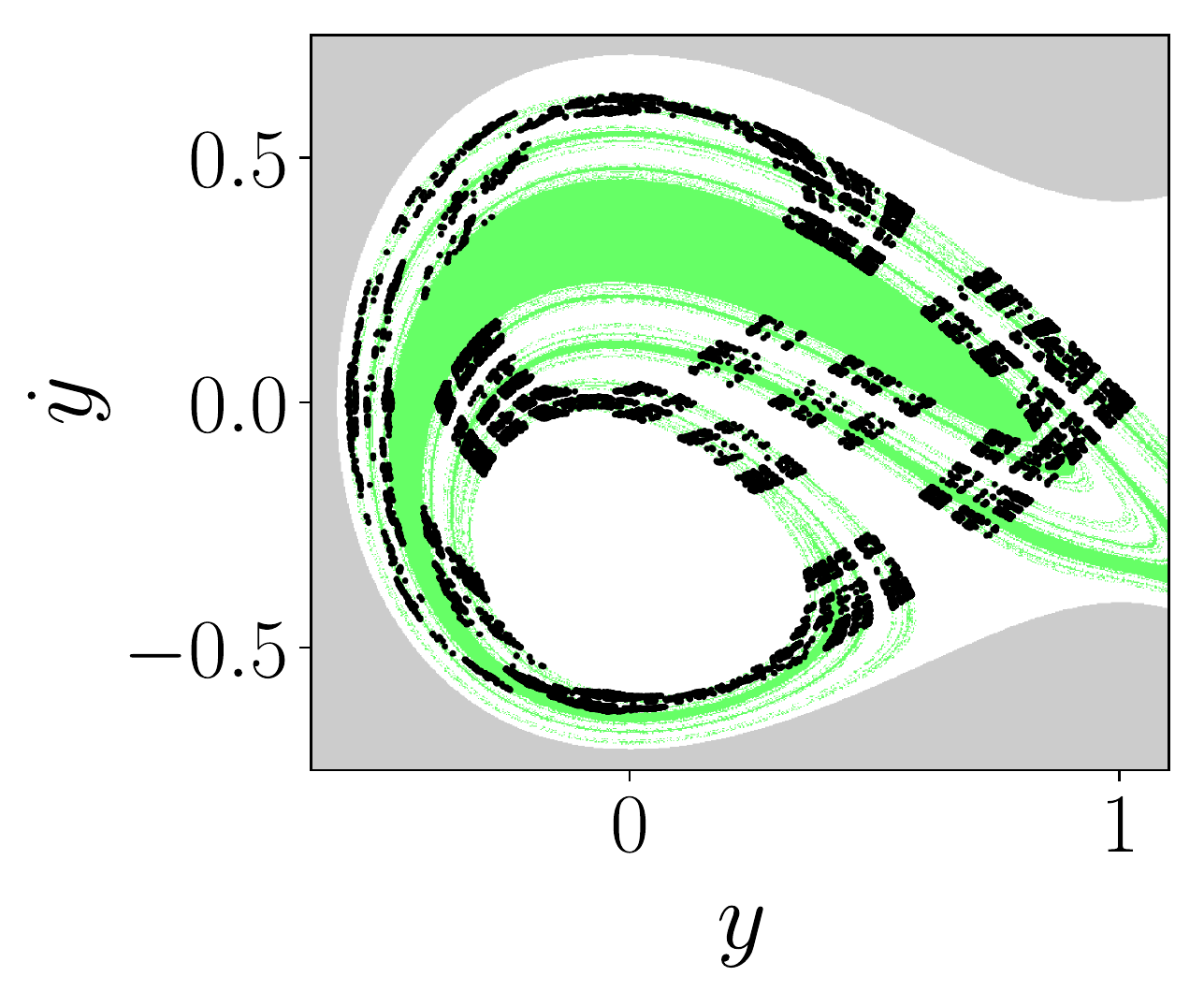}}
\subfigure[]{\includegraphics[width=0.35\textwidth]{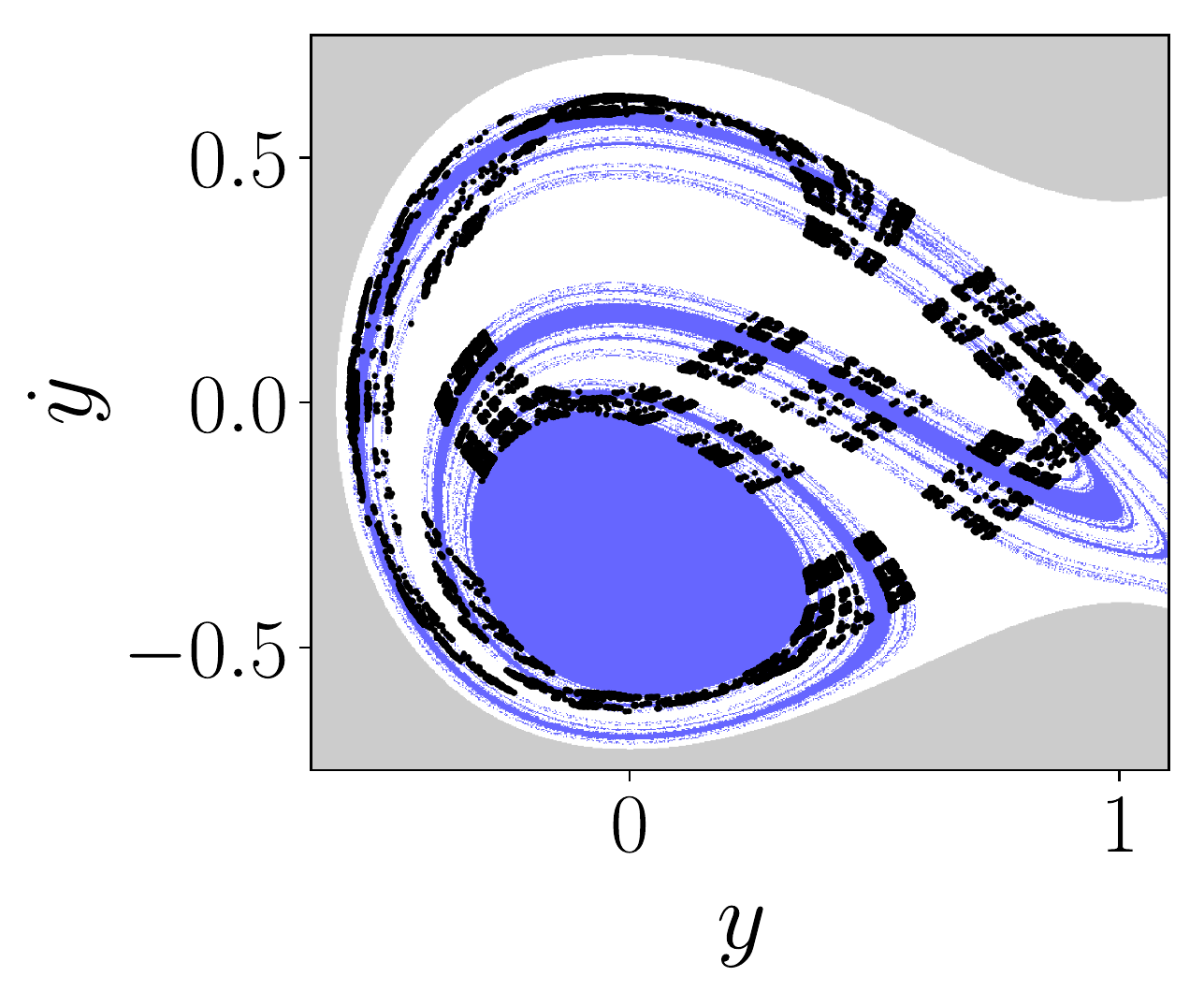}}
\end{center}
\caption{\label{hh_saddle_basins} \textbf{Saddle-straddle trajectories for the H\'enon-Heiles Hamiltonian $H=\frac{1}{2}(\dot{x}^2+\dot{y}^2)+\frac{1}{2}(x^2+y^2)+x^2y-\frac{1}{3}y^3$.} (a) Escape basins for the energy $E=0.25$. (b) Saddle-straddle trajectory for the red basin and the merged green and blue basins. (c)  Saddle-straddle trajectory for the green basin and (d) for the blue basin with other basins merged together.}
\end{figure*}

Our last example deals with a different system where the phase space does not contain attractors but the trajectories eventually diverge through a certain escape region. This is the Hénon-Heiles Hamiltonian given by $H=\frac{1}{2}(\dot{x}^2+\dot{y}^2)+\frac{1}{2}(x^2+y^2)+x^2y-\frac{1}{3}y^3$. This is a system with a four dimensional phase space that can be studied in the plane by means of a suitable Poincaré section. If we take energy values above the critical energy $E_c=1/6\simeq 0.166$, the Hamiltonian system opens up presenting three different exits depending on the initial conditions. The escape basins are obtained from the computation of the trajectories on a regular grid of initial conditions. The three basins share the same boundary and show the Wada property \cite{aguirre_wada_2001}, as can be guessed by Fig.~\ref{hh_saddle_basins} (a) for the energy value $E=0.25$. When an initial condition violates the energy requirement $E=0.25$ it belongs to the forbidden region painted in black.

We first proceed to the computation of the three possible saddles with the saddle-straddle algorithm. Again we shall name the chaotic sets with indices $r$, $g$, and $b$ associated to the escape basins in Fig.~\ref{hh_saddle_basins} (b), (c) and (d), respectively.

Our procedure for the computation of the Hausdorff distance between the saddles gives the following results from the comparison of $n_p=10000$ points in each set: $d_H(r,g) = 0.087$, $d_H(r,b) = 0.058$ and $d_H(b,g) = 0.085$. The H\'enon-Heiles Hamiltonian lacks any attractors. Nevertheless, we can observe that the Hausdorff distances are small compared to the diameter of the chaotic set $r$ measured as $d_s(r)=1.5$. In this case, we can also conclude that the escape basins of the H\'enon-Heiles Hamiltonian have the Wada property since every point in one chaotic saddle is within a small distance of the other computed chaotic saddles in phase space.

\section{Comparison of available methods to assert the Wada property}

As we have already mentioned, several new methods have been established for the detection of the Wada property in dynamical systems. We summarize their properties in Tab.~\ref{tab1} in order to compare their characteristics.

\begin{table*}
\begin{tabular}{|l|p{3cm}|l|c|p{6cm}|}
\hline
Name & Type of system & Dim. & Computation& What we need \\
    &                 &       & time &  \\

\hline
Nusse-Yorke method \cite{nusse_wada_1996} & ODEs Hamiltonians Maps & 2D & 1 & It requires a detailed knowledge of the basin and the boundaries (accessible unstable periodic orbit embedded in the basin boundary). \\
\hline
Grid method \cite{daza_testing_2015} & Any dynamical system & N-D & 100  & It requires the basins and the dynamical system to compute parts of the basin at a higher resolution. \\
\hline
Merging method \cite{daza_ascertaining_2018} & Any dynamical system & N-D & 0.01 & It needs to know the basins, but not the dynamical system.\\
\hline
Saddle-straddle method  & ODEs Hamiltonians Maps &  2D & 1 & It needs to know the dynamical system, but not the basins.\\
\hline
\end{tabular}
\caption{\label{tab1} Comparison of the principal procedures to test if a basin of attraction has the Wada property.}
\end{table*}

Each method has its own benefits and drawbacks in different categories. Depending on the problem we are dealing with, we might choose one method or another. If the basins can be computed easily, we might decide to use the Merging method or the Grid method. The first one is much faster, while the latter is more precise though slower. {The computation times appearing in Tab. \ref{tab1} are adimensional time magnitudes relative to the execution of the saddle-straddle method for the forced pendulum, which takes around one hour on a desktop PC. These are approximate values that will depend strongly on the problem as well as of the architecture of the computation. Also, part of the computation time cannot be evaluated directly, as for example the effort needed to compute a basin of attraction to apply the merging algorithm.} The Nusse-Yorke method can be used for systems possessing an accessible unstable periodic orbit embedded in the basin boundary. It is a reliable method, but it requires a detailed study of the dynamical system to detect the basin cells plus the computation of the unstable manifold associated to this orbit, which can be cumbersome in some cases, as discussed for example in \cite{daza_wada_2017, BHs}.

The saddle-straddle method is limited to connected Wada basins in two dimensions. In the case of disconnected Wada basins, we have two or more distinct disconnected fractal boundaries each one separating three or more basins. The algorithm can detect different chaotic saddles but it will fail to recognize them as Wada. It is however a powerful technique since we can test the Wada property only with the dynamics of the system. The basins of attraction are not needed and the accuracy of the test is only dependent on the length of the computed time series. Therefore, it is an excellent option to investigate the Wada property accurately and fast in two dimensional basins.

Concerning the meaning of term accuracy in the context of the saddle-straddle method, our experiments show that for two time series corresponding to the same chaotic attractor, the Hausdorff distance decreases as a power of the number of points. { As such, we can identify a unique boundary up to some fixed tolerance. This fact may be used to identify identical sets without the comparison with the diameter of a set. A quick sketch of the method would be to compare the Hausdorff distance of two sets, lets say $r$ and $g$, with the Hausdorff distance of a sample of these two sets strictly smaller, $r_1$ and $g_1$. If the distance $d_H(r,c)$ is systematically lower than $d_H(r_1,g_1)$ then we are in presence of a chaotic saddle.} Nonetheless, the distance may hit an inferior limit due to spurious points in the time series or due to the numerical precision of the variables.

It must also be noticed that the Hausdorff distance is not the only way to compare sets of points. There are computational techniques based on the correlation dimension between time series \cite{kantz1994quantifying}, or even standard statistical techniques to compare two distributions such as the chi-squared test \cite{press2007numerical}. Another simpler approach is just to find the closest points between the two sets. If these two points belong to a saddle and they are very close, it is likely that they belong to the same set. However, there could be a situation with two close saddles and an attractor between them. In this respect, the Hausdorff distance provides more information and a more reliable test, but it is important to remind that we are always constrained by the numerical precision of our algorithms.

Finally, we would like to draw some attention on the similarities between the Merging method and the saddle-straddle method to test for the Wada property in dynamical systems. As presented here, the saddle-straddle method relies on the Hausdorff distance to match different sets of points. This technique can be related to the fattening of a set. Imagine that we draw a small circle around each point of a saddle with radius $d_H$. The part covered with these circles is the fattened set of the saddle. If another set of points, such as a different saddle trajectory, can fit in this fattened set, then we claim that the two sets of points belong to the same chaotic saddle, up to a resolution defined by the fattening parameter. This fattening technique has been used successfuly to match fractal boundaries on a finite grid in \cite{daza_ascertaining_2018,BHs}. The Hausdorff distance measures the smallest fattening radius of a set that we need such that two sets of points can fit into this fattened region. In this context, the two methods share a common idea on how to compare different sets. However, this article focuses on the invariant subset of the boundary that has been obtained with a very accurate method while \cite{daza_ascertaining_2018} takes its data from a computed basin of attraction with finite precision, { the accuracy of the answer that gives the merging method is limited by the resolution of the basin of attraction.}

\section{Conclusions}

Proving the Wada property in dynamical systems may require different approaches adapted to the particularities of the problem under study. This article is based on the idea involving the invariance of the chaotic saddle through the merging of several basins for the Wada property to occur. The saddle-straddle algorithm computes $n$ points of the saddles between the merged basins. These sets of points are compared and if they coincide with some accuracy, we conclude that the chaotic saddle is unique and the basins have the Wada property. The precision of the algorithm depends directly on the length of the time-series obtained from the saddle-straddle algorithm. Therefore, the main contribution of the current work is to present a new method for the detection of Wada basins using a purely dynamical approach.

\section*{Acknowledgements}

This work was supported by the Spanish State Research Agency (AEI) and the European Regional Development Fund (FEDER) under Project No. FIS2016-76883-P.

%
\end{document}